\documentclass[doublecol]{epl2} 

\title{Circuit QED scheme for realization of the Lipkin-Meshkov-Glick model}
\shorttitle{Title} 

\author{Jonas Larson}

\institute{Department of Physics, Stockholm University, 
106 91 Stockholm, Sweden}

\pacs{42.50.Pq}{Cavity quantum electrodynamics}
\pacs{05.30.Rt}{Quantum Phase Transitions}
\pacs{42.50.Ct}{Quantum description of interaction of light and matter}

\abstract{We propose a scheme in which the Lipkin-Meshkov-Glick model is realized within a circuit QED system. An array of $N$ superconducting qubits interacts with a driven cavity mode. In the dispersive regime, the cavity mode is adiabatically eliminated generating an effective model for the qubits alone. The characteristic long-range order of the Lipkin-Meshkov-Glick model is here mediated by the cavity field. For a closed qubit system, the inherent second order phase transition of the qubits is reflected in the intensity of the output cavity field. In the broken symmetry phase, the many-body ground state is highly entangled. Relaxation of the qubits is analyzed within a mean-field treatment. The second order phase transition is lost, while new bistable regimes occur.  }

\begin{document}

\maketitle

\section{Introduction}
During the last two decades, systems involving single atoms or superconducting qubits interacting with a quantized mode of a high-$Q$ resonator have supported some of the finest experiments analyzing fundamental aspect of quantum mechanics. Among these cavity/circuit quantum electrodynamics (QED) experiments, it is especially worth mentioning; entanglement generation~\cite{entcav,wallraff3}, quantum-classical transition~\cite{qclass}, and verification of the graininess of the quantized electromagnetic field~\cite{cavgrain}. Very recently, a new branch of cavity/circuit QED including several atoms/qubits has emerged. Coherent coupling of Bose-Einstein condensate~\cite{cavityBEC} as well as coupling of few qubits~\cite{wallraff3,wallraff1,sillenpaa} to a single cavity mode have been experimentally demonstrated. These experiments paves the way for controlled studies of novel cooperative phenomena of many-body systems.    

Microscopic many-body systems are at the heart of a wide range of physical branches, e.g. condensed matter, quantum information processing, Bose-Einstein condensations, and ultracold atoms. With experimental progress, much attention has been especially paid to phase transitions and more recently to ground state entanglement properties. At zero temperature, a phase transition in these systems is driven by quantum fluctuations and is thereby termed  {\it quantum phase transition}~\cite{sachdev}. In the realm of quantum information processing, determining the quantum properties of the many-body ground state is of central significance and especially the amount of entanglement distributed among the entities~\cite{manyent}. Spin models, describing interaction between spin 1/2 particles are of special importance as each spin can be ascribed a qubit state. Most common are lattice models, e.g. {\it Ising} and $XX$ models~\cite{sachdev}, characterized by short or semi-short range interaction. Lately, also long range models like the {\it Lipkin-Meshkov-Glick} (LMG) have gained extra recognition since they might be realizable in the near future in various systems. 

The LMG model has its origin in nuclear physics~\cite{lmg}. In the recent past, several proposals for realization of the LMG model in different contexts have been put forward, for example within trapped ions~\cite{lmgtrap}, Bose-Einstein condensates in double-wells~\cite{lmgbecdw} or in cavities~\cite{lmgbeccavity}, or of non-interacting atoms/qubits within a cavity~\cite{parkins,tsomokos}. The general properties of the LMG model have been considered by numerous authors, focusing on entanglement properties~\cite{lmgent} and finite size effects~\cite{lmgfinite}. 

In this paper we show how the symmetric LMG model may be achieved by extending the recent experiment of ref.~\cite{wallraff1}; an array of superconducting qubits are identically coupled to a cavity mode. The resonator is externally pumped, and in the dispersive regime the cavity field is adiabatically eliminated yielding an effective model for the qubits. The quantized cavity mode serves as a {\it quantum bus}~\cite{wallraff3}, which induces the long range interaction between the qubits. For zero coupling of the qubits to its environment, we consider the full many-body problem and calculate the ground state entanglement both for the LMG model and an extended model also realizable in the same setup. Decoherence of the qubits is analyzed within a mean-field approach. We demonstrate that bistable dynamics emerges in this situation, while the second order quantum phase transition of the LMG model is lost. All these ground-state properties are accessible via detection of the output cavity field leaking out of the end mirrors.

\section{Model system}
Inspired by the recent circuit QED experiments~\cite{wallraff3,wallraff1,sillenpaa}, and the quest for engineering multi-qubit systems, we consider an array of $N$ superconducting qubits coupled to a single mode of an optical resonator. The individual effective qubit-field coupling strength $g_i$ ($i=1,\,2,\,...,N$) are taken to be all equal, i.e. $g_i\equiv g$, $\forall\,\,i$. This is in agreement with ref.~\cite{wallraff1} where the fluctuations of the $g_i$'s were only a few percent. The quality factor $Q$ of the resonator is assumed large enough to support well resolved mode frequencies, but still not infinite such that the photon decay rate $\kappa\neq0$. The cavity is externally driven via one side mirror by a pump with frequency $\omega_p$ and amplitude $\eta$. Considering an interaction frame rotating with the pump frequency and introducing the cavity-pump detuning $\Delta_c=\omega_c-\omega_p$ and the atom-pump detuning $\Delta_a=\omega_a-\omega_p$, the Hamiltonian is written
\begin{equation}
\label{ham1}
\begin{array}{lll}
\hat{H}_D & = & \hat{H}_0+\hat{\mathcal{H}}_{cav}+\hat{\mathcal{H}}_{qu}\\ \\
& & \hbar\Delta_c\hat{a}^\dagger\hat{a}-i\hbar\eta\left(\hat{a}-\hat{a}^\dagger\right)+\hbar\Delta_a\hat{S}_z\\ \\ & & +g\left(\hat{a}+\hat{a}^\dagger\right)\hat{S}_x
+\hat{\mathcal{H}}_{cav}+\hat{\mathcal{H}}_{qu}.
\end{array}
\end{equation}
Here, $\hat{a}$ and $\hat{a}^\dagger$ are annihilation and creation boson operators of the cavity mode respectively, $\hat{S}_\alpha=\frac{1}{2}\sum_{i=1}^N\hat{\sigma}_\alpha^{(i)}$ is the collective spin operator of component $\alpha$ ($=x,\,y,\,z$) with $\hat{\sigma}_x^{(i)}=|e\rangle_i\,_i\!\langle g|+|g\rangle_i\,_i\!\langle e|$ and $\hat{\sigma}_z^{(i)}=|e\rangle_i\,_i\!\langle e|-|g\rangle_i\,_i\!\langle g|$ the regular Pauli matrices acting on the ground $|g\rangle$ and excited $|e\rangle$ state of qubit $i$, and finally $\hat{\mathcal{H}}_{cav}$ and $\hat{\mathcal{H}}_{qu}$ represent coupling to surrounding reservoirs leading to respectively photon leakage and qubit decoherence. 

In the absence of losses and pumping, $\hat{H}_D$ is the Dicke Hamiltonian introduced to describe superradiance of $N$ two-level atoms interacting with a quantized cavity mode. If $g<\sqrt{\omega_a\omega_c}$, the ground state of the system is said to be in the {\it normal phase} charactericed by vacuum cavity field and all qubits in their ground states, while if $g>\sqrt{\omega_a\omega_c}$ the ground state is an entangled {\it superradiance phase} where the field is in a Schr\"odinger cat and the qubits partly excited~\cite{dicke}. The quantum phase transition between the two phases is of second order nature without the rotating wave approximation. The qubit-field coupling is inversely proportional to the square root of the effective mode volume $V$, implying that if the density $\rho_0=N/V$ is fixed we may write $g=g_0/\sqrt{N}$. The scaling of the coupling gives a well defined thermodynamic limit. In the presence of external pumping, one must assume $\eta=\eta_0\sqrt{N}$ in order to have a proper thermodynamic limit. Note that this is intuitive, if the cavity volume is enlarged the pumping should be increased accordingly.

If qubit relaxation $\gamma$ and qubit dissipation $\gamma'$ can be set to zero within the time-scales of interest, the system state $\hat{\rho}$ obeys the master equation
\begin{equation}
\label{master}
\frac{d}{dt}\hat{\rho}=\frac{i}{\hbar}[\hat{\rho},\hat{H}_0]+\hbar\kappa\left(2\hat{a}\hat{\rho}\hat{a}^\dagger-\hat{a}^\dagger\hat{a}\hat{\rho}-\hat{\rho}\hat{a}^\dagger\hat{a}\right).
\end{equation}
Heisenberg equations of motion become
\begin{equation}
\label{eom}
\begin{array}{l}
\displaystyle{\dot{\hat{a}}=-(\kappa+i\Delta_c)\hat{a}+\eta_0\sqrt{N}-i\frac{g_0}{\sqrt{N}}\hat{S}_x+\sqrt{2\kappa}\hat{a}_{in}(t)},\\ \\
\displaystyle{\dot{\hat{S}}_x=-\Delta_a\hat{S}_y+\frac{g_0}{\sqrt{N}}\left(\hat{a}+\hat{a}^\dagger\right)\hat{S}_z},\\ \\
\displaystyle{\dot{\hat{S}}_y=\Delta_a\hat{S}_x+i\frac{g_0}{\sqrt{N}}\left(\hat{a}-\hat{a}^\dagger\right)\hat{S}_z},\\ \\
\displaystyle{\dot{\hat{S}}_z=-i\frac{g_0}{\sqrt{N}}\left(\hat{a}-\hat{a}^\dagger\right)\hat{S}_y-\frac{g_0}{\sqrt{N}}\left(\hat{a}+\hat{a}^\dagger\right)\hat{S}_x}.
\end{array}
\end{equation}
Here, dot represent time-derivative and $\hat{a}_{in}(t)$ is the input Langevin noise obeying $\langle\hat{a}_{in}(t)\rangle=\langle\hat{a}_{in}^\dagger(t)\hat{a}_{in}(t')\rangle=0$ and $\langle\hat{a}_{in}(t)\hat{a}_{in}^\dagger(t')\rangle=\delta(t-t')$~\cite{zoller}. In what follows, the Langevin force $\hat{a}_{in}(t)$ will be omitted. In order to derive an effective spin Hamiltonian we assume a separation of time-scales, the cavity field is taken to evolve on a much shorter time scale than the qubits. This implies $|\kappa+i\Delta_c|\gg g_0$, and in this regime we may consider the steady state of the cavity field
\begin{equation}
\label{ss}
\hat{a}_{ss}=\frac{\eta_0\sqrt{N}-ig_0\hat{S}_x/\sqrt{N}}{\kappa+i\Delta_c}.
\end{equation}
Inserting this expression, together with its corresponding hermite conjugate, into the equations of motion for the collective spin operators the resulting equations defines an effective Hamiltonian for the qubits. To order $N^0$ one obtains
\begin{equation}
\label{effham}
\hat{H}_{eff}=\hbar\Delta_a\hat{S}_z-\hbar h\hat{S}_x-\frac{\hbar\lambda}{N}\left(\hat{S}_y^2+\hat{S}_z^2\right),
\end{equation}
where
\begin{equation}
\begin{array}{lll}
\displaystyle{h=-\frac{2g_0\eta_0\kappa}{\kappa^2+\Delta_c^2}}, & & \displaystyle{\lambda=\frac{g_0^2\Delta_c}{\kappa^2+\Delta_c^2}}.
\end{array}
\end{equation}
For driving at resonance with the atomic transition, $\Delta_a=0$, the above Hamiltonian is unitary equivalent with the isotropic LMG model via $U=\exp\left[-i\frac{2\pi}{3}\left(\hat{S}_x+\hat{S}_y+\hat{S}_z\right)/\sqrt{3}\right]$. The long-range coupling, inherent in the squares of the collective spin operators, is a direct outcome of the cooperative qubit-field interaction. The second term, proportional to $\hat{S}_x$, characterizes driving of the qubits, and derives from the driving of the cavity field. Such a term is expected since driving of the qubits or field is unitary equivalent in the Dicke model. We note that a special case of the LMG model has been discussed in the context of circuit QED earlier, with the focus on few qubits and a closed system~\cite{tsomokos} and not on phase transitions nor bistability.

\section{Many-body properties of the effective qubit system}
The isotropic LMG model, i.e. $\Delta_a=0$ in eq.~(\ref{effham}), is analytically solvable. Since the total spin $\hat{\mathbf{S}}^2$ commutes with $\hat{H}_{eff}$ and so does $\hat{S}_x$, the eigenstates of $\hat{H}_{eff}$ are $|S,M\rangle_x$, where $\hat{\mathbf{S}}^2|S,M\rangle_x=S(S+1)|S,M\rangle_x$ and $\hat{S}_x|S,M\rangle=M|S,M\rangle_x$. The eigen energies per particle
\begin{equation}
\frac{E(S,M)}{N}=-\frac{\hbar h}{N}M-\frac{\lambda}{N^2}\left(S(S+1)-M^2\right),
\end{equation}
where the ground state is to be found in the maximum spin sector, $S=N/2$, and 
\begin{equation}\label{mnum}
M_0=\left\{\begin{array}{lll}
I(hN/2), & & |h|<\lambda\\ \\
N/2, & & |h|\geq\lambda.
\end{array}\right.
\end{equation}
Here, $I(x)$ is the integer part of $x$ rounded off to nearest integer. Without loss of generality, we have restricted the analysis to the ferromagnetic case $\lambda>0$. $\langle\hat{S}_x\rangle=M_0$ being an order parameter, the ground state possess a second order phase transition at $h_c=\pm\lambda$. For $|h|\geq\lambda$ the ground state is ferromagnetic, while in the broken symmetry phase $|h|<\lambda$ the system exhibits an multi-partite entangled ground state.

For a symmetric multi-partite qubit state, the reduced density matrix $\hat{\rho}_{2q}$ for two of the qubits is easily obtainable~\cite{moller}. The {\it concurrence} is a proper measure of entanglement for a general, mixed or pure, two qubit state. Given $\hat{\rho}_{2q}$, we introduce the operator $\hat{\varrho}_{2q}=\hat{\rho}_{2q}(\hat{\sigma}_{1y}\otimes\hat{\sigma}_{2y})\hat{\rho}_{2q}^*(\hat{\sigma}_{1y}\otimes\hat{\sigma}_{2y})$ where $\hat{\sigma}_{iy}$ is the Pauli $y$-matrix acting on qubit $i$ and asterix denotes complex conjugation. The concurrence is defined as
\begin{equation}
\label{conc}
\mathcal{C}=\mathrm{max}\left\{0,\mu_1-\mu_2-\mu_3-\mu_4\right\},
\end{equation}
with $\mu_j$ the square root of the $j$'th eigenvalue of $\hat{\varrho}_{2q}$ ordered such that $\mu_1\geq\mu_2\geq\mu_3\geq\mu_4$. For a disentangled state $\mathcal{C}=0$, while $\mathcal{C}=1$ characterices a maximally entangled one. In the case of the isotropic LMG model one has the rescaled concurrence, $C_R=(N+1)\mathcal{C}$, given by~\cite{moller}
\begin{equation}
\label{conc2}
\begin{array}{lll}
C_R & = & \displaystyle{\frac{1}{2N}}\left\{N^2-4M^2-\right.\\
& & \left.\sqrt{(N^2-4M^2)[(N-2)^2-4M^2]}\right\}.
\end{array}
\end{equation}
  
\begin{figure}
\includegraphics[width=8cm]{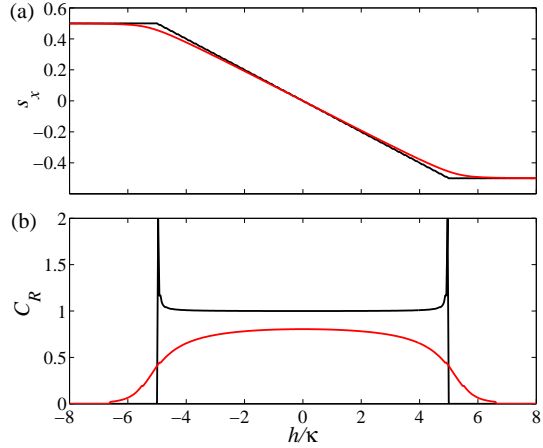}
\caption{(Colour online) The order parameter $s_x=\langle\hat{S}_x\rangle/N$ (a) and scaled concurrence $C_R$ (b). For the red line $\Delta_a=\kappa/5$, while for the black one $\Delta_a=0$. It follows that the phase transition at $|h|=\lambda_c=5\kappa$ is lost once $\Delta_a\neq0$. The parameters are $g_0=100\kappa$ and $\Delta_c=2000\kappa$.  }
\label{fig1}
\end{figure}
  
For our numerical results we try to employ physically relevant parameter values. Experimentally, the photon decay rate $\kappa$ is of the order 1 to 50 MHz~\cite{wallraff1,wallraff2}. The qubit-field coupling $g_0\sim600$ MHz in the strong coupling regime experiments~\cite{wallraff1,wallraff2}. In the following we will express frequencies in units of $\kappa$, and take $g_0=100\kappa$. Furthermore, to assure validity of the adiabatic elimination of the cavity field, we use $\Delta_c=2000\kappa$, i.e. 20 times larger than the qubit-field coupling. In fig.~\ref{fig1}, the order parameter $s_x\equiv\langle\hat{S}_x\rangle/N$ and scaled concurrence $C_R$ are shown as function of $h/\kappa$. Varying of $h$ is preferably achieved by tuning the pump amplitude $\eta_0$ since $h\propto\eta_0$. In the two plots, the black lines depict results for the LMG, i.e. $\Delta_a=0$, while the red lines give the same for $\Delta_a=\kappa/5$. The number $N$ of qubits is here 200, but the results are relatively insensitive to it. That is, even for few qubits is the scheme efficient for entanglement generation. We note that the second order phase transition at $|h|=\lambda_c=5\kappa$ vanishes for non-zero $\Delta_a$. For the LMG model, the rescaled concurrence is approximately 1 in the broken symmetry phase indicating maximal distribution of the entanglement between the qubits. In the vicinity of the critical points, the concurrence increases abruptly, hence giving a clear signal of the phase transition. The amount of quantum correlation for $|h|<\lambda_c$ is lower when $\Delta_a\neq0$, but contrary to the LMG model, the concurrence survives also for regimes with $|h|>\lambda_c$. However, at $h\approx6.5\kappa$ the entanglement suddenly vanishes. This rapid disappearance is only present in the concurrence and not in the order parameter $s_x$, and is therefore not an outcome of a phase transition. Indeed, {\it sudden death of entanglement} has been studied in great detail lately~\cite{entsud}.  

Apart from constituting an effective qubit-qubit interaction, the quantum properties of the cavity field allow for non-destructive detection of the multi-partite qubit state. The amount of leakage of photons out of the end mirror is directly proportional to the photon number in the resonator~\cite{nondes}. Since,
\begin{equation}
n_{ss}=\frac{\eta_0^2N}{\kappa^2+\Delta_c^2}+\frac{g_0^2}{\kappa^2+\Delta_c^2}\frac{\langle\hat{S}_x^2\rangle}{N},
\end{equation}
measuring $n_{ss}$ gives a handle of the qubit state. In particular, in the LMG case we have $\langle\hat{S}_x^2\rangle=M_0^2$ for the ground state with $M_0$ given in eq.~(\ref{mnum}).

\section{Influence of qubit decoherence}
In the case of two-photon Raman coupling via a highly detuned third level, the two qubit states $|g\rangle$ and $|e\rangle$ can be made insensitive to decay. However, for a direct transition, the excited $|e\rangle$ state will decay with some rate $\gamma$. Such irreversible processes will affect the system properties. The collective decoherence and dissipation are modeled with a master equation for the qubits,
\begin{equation}
\begin{array}{lll}
\displaystyle{\frac{d}{dt}\hat{\rho}_q} &  = & \displaystyle{\frac{i}{\hbar}[\hat{H}_{eff},\hat{\rho}_q]}\\ \\
& & +\displaystyle{\frac{\gamma}{N}}\left(2\hat{S}_+\hat{\rho}_q\hat{S}_--\hat{S}_-\hat{S}_+\hat{\rho}_q-\hat{\rho}_q\hat{S}_-\hat{S}_+\right)\\ \\
& & +\displaystyle{\frac{\gamma'}{N}}\left(2\hat{S}_x\hat{\rho}_q\hat{S}_x-\hat{S}_x^2\hat{\rho}_q-\hat{\rho}_q\hat{S}_x^2\right),
\end{array}
\end{equation}
where $\hat{S}_\pm=\hat{S}_x\pm\hat{S}_y$ and $\gamma'$ characterizes the decoherence rate. Hence, the $\gamma$-term describes decay of the excited qubit level and the $\gamma'$-term loss of coherence due to interaction with its environment. We will assume the thermodynamic limit, where we can neglect quantum fluctuations and thereby factorize $\langle\hat{A}\hat{B}\rangle=\langle\hat{A}\rangle\langle\hat{B}\rangle$ for any operators $\hat{A}$ and $\hat{B}$. Hence, we consider a mean-field approach. Denoting $s_\alpha\equiv\langle\hat{S}_\alpha\rangle/N$, $\alpha=x,\,y,\,z$, we obtain the following semiclassical equations of motion
\begin{equation}\label{heis}
\begin{array}{l}
\dot{s}_x=-\Delta_as_y-\gamma s_zs_x,\\ \\
\dot{s}_y=\Delta_as_x+hs_z-\lambda s_zs_x-\gamma s_zs_y,\\ \\
\dot{s}_z=-hs_y+\lambda s_ys_x+\gamma\left(s_x^2+s_y^2\right)
\end{array}
\end{equation}
in the thermodynamic limit. In (\ref{heis}), only terms of order $N^0$ are kept, and as a consequence contribution from the dissipation vanishes. Throughout this section we will set $\Delta_a=0$. In addition to (\ref{heis}), we have the angular momentum conservation constrain $s_x^2+s_y^2+s_z^2=1$. The system of equations (\ref{heis}) posses in general several fixed points $(\bar{s}_x,\bar{s}_y,\bar{s}_z)$. Analyzing the fixed points, i.e. study the steady state solutions ($\dot{s}_x=\dot{s}_y=\dot{s}_z=0$) of (\ref{heis}), one finds that if $\gamma=0$ the many-body results of the previous section are regained. If, on the other hand, $\gamma\neq0$ it directly follows that either $\bar{s}_x$ or $\bar{s}_z$ must be identically zero. 

\begin{figure}
\includegraphics[width=8cm]{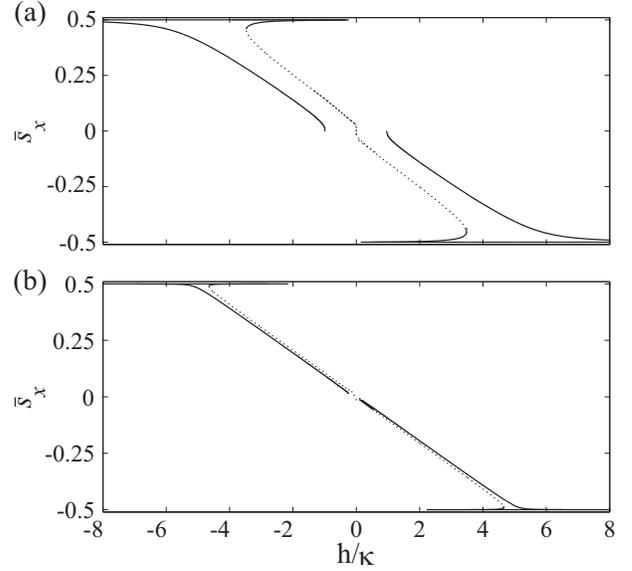}
\caption{ The steady state solutions $\bar{s}_x$ of the truncated Heisenberg equation (\ref{heis2}). The solid lines are stable solutions, while dotted lines represent unstable solutions. The parameters are $g_0=100\kappa$, $\Delta_c=2000\kappa$, and $\gamma=\kappa/5$ (a) or $\gamma=\kappa/50$ (b).  }
\label{fig2}
\end{figure}

For $\bar{s}_x=0$, the steady state solution is easily found as
\begin{equation}\label{heis3}
\begin{array}{lll}
\displaystyle{\bar{s}_y=\frac{h}{\gamma}}, & & 
\displaystyle{\bar{s}_z=\sqrt{1-\left(\frac{h}{\gamma}\right)^2}},
\end{array}
\end{equation}
valid for $|h|<\gamma$. In the other case, $\bar{s}_z=0$, the steady state solutions obey 
\begin{equation}\label{heis2}
\begin{array}{lll}
\displaystyle{\bar{s}_x=\sqrt{1-\bar{s}_y^2}},\\ \\
\displaystyle{\bar{s}_y^4-\left[1-\left(\frac{h}{\lambda}\right)^2\right]\bar{s}_y^2-\frac{h\gamma}{\lambda}\bar{s}_y+\left(\frac{\gamma}{\lambda}\right)^2=0}.
\end{array}
\end{equation}
This set of solutions shows an {\it imperfect transcritical bifurcation} pattern, see~\cite{nonlin}. The last constant term $(\gamma/\lambda)^2$ cause the bifurcation to become imperfect. As an outcome, the phase transition encountered in the lossless case ($\gamma=0$) at $|h|=\lambda_c$ is lost and the transition is instead smooth. In addition, multiple solutions may exist of which some are stable and others unstable. The stability is found via a {\it linear stability analysis} around the fixed points~\cite{nonlin}, i.e. perturb around the steady state solutions and check whether the perturbations increase or decrease with time. In fig.~\ref{fig2} we display the steady state solution $\bar{s}_x$ corresponding to the solutions (\ref{heis2}) for $\gamma=\kappa/5$ in (a) and $\gamma=\kappa/50$ in (b). The value of $\gamma$ in (a) gives a realistic decay rate of an excited state in a qubit~\cite{wallraff2}. In the figure, solid lines represent stable solutions, while the unstable solutions are marked by dotted lines. 

It is seen in fig.~\ref{fig2} that a maximum of three stable solutions may coexist given a set of parameters. The one solution picked by the system is determined by details of the preparation and dynamical process taken by the experiment at hand. For example, by dynamically varying a system parameter the evolution must not imply that the system follows the ground state. However, if the characteristic time scale of the evolution is long compared to the inherent time scale for quantum fluctuations it is likely that transitions from stable solutions with higher energy will decay into stable solutions with a lower energy. Of the three stable solutions of fig.~\ref{fig2}, the solution with smallest $|\bar{s}_x|$ is presumably having the lowest energy. This assumption derives from estimating the energy utilizing the energy density $\mathcal{E}=-\hbar h\bar{s}_x-\hbar\lambda\left(\bar{s}_y^2+\bar{s}_z^2\right)$. Thus, the coupling of the qubit to the environment enters in the actual value of the mean-field values $(\bar{s}_x,\bar{s}_y,\bar{s}_z)$ and not directly in the energy density. When the solid lines terminate at $|h|\approx\kappa$ of fig.~\ref{fig2} (a), the corresponding solutions become complex. It therefore follows that there may occur parameter regions in which no stable solution exists. The solution (\ref{heis3}), valid for $|h|<\gamma$, is stable but still its existence does not warrant stable solutions within all parameter regimes. Thus, tuning $h$ across $|h|\approx\kappa$ from $|h|>\kappa$, and assuming the system to follow adiabatically its instantaneous ground state, will cause sudden jumps in the physical quantities. This bistability will be reminiscent of a first order dynamical phase transition. It should be pointed out that these results are derived in the thermodynamic limit, hence it does not necessarily imply that the finite size system also lacks stable solutions within these parameters. As discussed in the previous section, the output cavity field being proportional to $n_{ss}$ contains information about the actual state of the qubits. This, of course, holds also in this mean-field treatment and thereby one should be able to detect the dynamical bistability indicated by fig.~\ref{fig2}. The second (b) plot makes clear that in the limit of no qubit decay, $\gamma=0$, the results of the previous section is captured. We note that optical bistability has gained renewed interest lately in optomechanical systems~\cite{jonas}.  

\section{Conclusion}
We have shown that under proper conditions, the LMG many-body spin model can be realized within circuit QED by considering an array of superconducting qubits coupled to one quantized mode of a high-$Q$ resonator. The symmetric LMG model is obtained when the cavity pump is resonant with the qubit transition, and it was demonstrated how the cavity output field reveals sufficient information for determining the phase of the many-body qubit ground state, i.e. the ferromagnetic or the broken symmetric one. The many-body entanglement properties of the broken symmetric phase was discussed. Driving the resonator off resonant with respect to both the cavity and the qubits, the phase transition of the LMG model is lost and the ground state becomes in general entangled also in the otherwise ferromagnetic phase. 

Within a mean-field treatment we considered the situation where qubit relaxation and dephasing are taken into account. An imperfect bifurcation structure emerges where the second order phase transition of the LMG model vanishes, but bistable properties appear. Driving the system through such bistable points would probably lead to sudden changes in the measured quantities.

In our treatment, all qubits were assumed identical and also to couple identically to the cavity mode. In ref.~\cite{wallraff1}, an experiment containing three qubits was manufactured and it was found that the fluctuations of the couplings were only a few percent. For larger number of qubits and greater fluctuations, new phenomena is likely to occur. Of particular interest is the realization of spin glasses. Spin glasses are normally outcomes of disorder within spin models~\cite{glass}. Thus, the considered model might be suitable for controlled generation of spin glass systems. An advantage of considering superconducting qubits rather than ultracold atoms is the absence of particle motion. It is known that the additional degrees of freedom induced by particle motion can greatly affect the system dynamics and its phase diagram properties~\cite{jonas2}.

\acknowledgments
The author acknowledges financial support from the Swedish research council--Vetenskapsr\aa det/VR.

\end{document}